\documentstyle[12pt]{article}

\begin{document}
\begin{flushright}
  RIMS 934\\
  July 1993\\
  hep-th/9307115
\end{flushright}
\begin{center}
\begin{Large}
{\bf Topological String, Matrix Integral,}\\[3mm]
{\bf and Singularity Theory}${}^\star$
\end{Large}

\vspace{25pt}

\noindent
T. Nakatsu,\raisebox{2mm}{{\footnotesize 1}{$\dagger$}}
A. Kato,\raisebox{2mm}{{\footnotesize 2}}
M. Noumi\raisebox{2mm}{{\footnotesize 2}} and
T. Takebe\raisebox{2mm}{{\footnotesize 3}}

\vspace{18pt}
\begin{small}
$~^{1}${\it Reserch Institute for Mathemtical Sciences,
Kyoto University} \\
{\it Sakyo-ku, Kyoto 606, Japan}

$~^{2}${\it Department of Mathematical Sciences, University of Tokyo} \\
{\it Komaba, Meguro-ku, Tokyo 153, Japan} \\

$~^{3}${\it Department of Mathematical Sciences, University of Tokyo} \\
{\it Hongo, Bunkyo-ku, Tokyo 113, Japan} \\

\end{small}

\vspace{25pt}

\underline{ABSTRACT}
\end{center}

\vspace{10pt}
\begin{small}
We study the relation between topological string theory and
singularity theory using the partition function of $A_{N-1}$
topological string defined by matrix integral of Kontsevich type.
Genus expansion of the free energy is considered, and the genus $g=0$
contribution is shown to be described by a special solution of
$N$-reduced dispersionless KP system. We show a universal
correspondences between the time variables of dispersionless KP
hierarchy and the flat coordinates associated with versal deformations
of simple singularities of type $A$. We also study the behavior of
topological matter theory on the sphere in a topological gravity
background, to clarify the role of the topological string in the
singularity theory. Finally we make some comment on gravitational
phase transition.
\end{small}

\vfill
\hrule

\vskip 3mm
\begin{small}

\noindent{$\star$} Partly supported by the Grant-in-Aid
for Scientific Research
from the Ministry of Education, Science and Culture
(No.04-2597, No.05854014, No.05740083).

\noindent{$\dagger$} A Fellow of the Japan
Society of the Promotion of Science for Japanese Junior Scientists.

\end{small}

\newpage

Recently much attentions have been paid to the understanding of
topological strings in terms of Gauss-Manin systems
associated with versal deformations of simple singularities
\cite{Lossev},\cite{EKYY}.
At present, however, it is not yet clear what is the interrelation
between the Gauss-Manin structure and the integrable hierarchy
\cite{Kon},\cite{Morozov} which also plays an important role in
topological strings.

In this letter, by using a more fundamental framework, we describe the
relation between topological string and singularity theory. We study
$A_{N-1}$ topological string which is defined by matrix integral of
Kontsevich type. By splitting its free energy into the contribution
of each genus, the genus $g=0$ contribution is shown to be described
by a special solution of $N$-reduced dispersionless KP system. We
also establish a universal correspondences between the time variables
of dispersionless KP hierarchy and the flat coordinates associated
with versal deformations of simple singularities of type $A$. The
combination of these two results clarifies the relation between the
$A_{N-1}$ topological string and a simple singularity of type
$A_{N-1}$. In order to search the role of the topological string in
the singularity theory, we study the behavior of the gravitational
primary fields on the sphere ($g=0$)in a fixed topological gravity
background. The possibility of gravitational phase transition of the
matter theory is strongly suggested.

In this letter we will describe our results briefly.
Their derivations will be given in a separate publication
\cite{noumiseminar}.

\vskip 3mm

Let us begin by some conceptual observations on $A_{N-1}$ topological
string. The $A_{N-1}$ topological string is the coupled system of
$A_{N-1}$ topological minimal matter and topological gravity
\cite{witten1}. The physical observables are denoted by
$\sigma_{nN+m}$ ( $ n \in {\bf Z}_{\geq 0}, 1 \leq m \leq N-1$).
$\sigma_1=P,\sigma_2,\cdots,\sigma_{N-1}$ are gravitational primary
fields are $\sigma_{nN+m}$ are the $n$-th gravitational descendants
of $\sigma_m$.

The free energy ${\cal F}$ of the $A_{N-1}$ topological string
should be given by;
\begin{equation}
  {\cal F}(\{t_{k}\}) = \sum_{ \{k_i\}, \{d_i\}, s }
  \langle
  \sigma_{k_{1}}^{d_{1}}
  \cdots
  \sigma_{k_{s}}^{d_{s}}
  \rangle
  \frac{ t_{k_{1}}^{d_{1}} \cdots t_{k_{s}}^{d_{s}}}
  {d_{1}! \cdots d_{s}! },
\label{Ftotal}
\end{equation}
where the correlation function
$
  \langle
  \sigma_{k_{1}}^{d_{1}}
  \cdots
  \sigma_{k_{s}}^{d_{s}}
  \rangle
$
should be described by the intersection theory on the compactified
moduli space $\overline{\cal M}_{g,s}$ of Riemann surface $\Sigma$ of
genus $g$ with $s$ punctures \cite{witten1}, where $g$ is fixed by the
relation:
\begin{equation}
  2(g-1)(N+1) = \sum_{i=1}^{s} d_i \{ k_i - (N+1) \}.
  \label{genus}
\end{equation}
This is nothing but the ghost number conservation law for the system.

Now we define the free energy ${\cal F}(\{ t_{k} \})$ in eq (\ref{Ftotal})
by the following matrix integral of Kontsevich type:
\begin{eqnarray}
  e^{{{\cal F}\left(\{ t_{k} \}\right)}}
  & := & \lim_{M \rightarrow \infty} Z_{N}^{(M)} (\Lambda),
 \label{defF} \\
   Z_{N}^{(M)} (\Lambda) & :=&
 \frac{
  \displaystyle
  \int dX \exp \left[- {\rm Tr} \left\{
   \frac{(X+\Lambda)^{N+1}}{N+1}
   - \frac{\Lambda^{N+1}}{N+1}
   + \Lambda^N X
   \right\}\right]
  }
  {
  \displaystyle
  \int dX \exp \left[ - \sum_{k=0}^{N-1} {\rm Tr} \left\{
   \Lambda^{k} X
   \Lambda^{N-1-k} X
   \right\} \right]
  },
\label{GKM}
\end{eqnarray}
where $X$ is a $M \times M$ hermitian matrix and $\Lambda = {\rm
diag~} (\lambda_1,\cdots,\lambda_M)$. The parameters $t_k$ are
related with $\Lambda$ by $t_k=\frac{1}{k} {\rm Tr} \Lambda^{-k}$,
which can be treated as independent
variables when $M$ is sufficiently large. The matrix integral in
(\ref{GKM}) is studied in \cite{Kon} ($N=2$ case) and in
\cite{Morozov} ($N\geq2$ case). Especially they showed that $e^{{\cal F}}$
is a $\tau$-function of $N$-reduced KP system and that, after an
appropriate shift of the marginal parameter $t_{N+1}$, it satisfies
the following Virasoro condition \cite{DVV} ;
\begin{eqnarray}
 0 &=& {\cal L}_n e^{{{\cal F}\left(\{ t_{k} \}\right)}} \nonumber \\
 &=&
 \sum_{k\geq 1} k t_{k}
 \frac{\partial}{\partial t_{k+Nn}} e^{{{\cal F}\left(\{ t_{k} \}\right)}}
 + \frac{1}{2} \sum_{k+l=nN}
 \frac{\partial}{\partial t_{k}}
 \frac{\partial}{\partial t_{l}} e^{{{\cal F}\left(\{ t_{k} \}\right)}}
 \nonumber \\
 && +
 \frac{1}{2} \sum_{k+l=N} k l
  t_{k}~ t_{l} e^{{{\cal F}\left(\{ t_{k} \}\right)}}
 \delta_{n+1,0}
 + \frac{N^2 - 1}{24} e^{{{\cal F}\left(\{ t_{k} \}\right)}} \delta_{n,0}
 \label{Virasoro}
\end{eqnarray}
where $n\geq -1$.

Let us combine the above result with our first observation. By
applying the relation (\ref{genus}) to (\ref{Ftotal}) we can split
${{\cal F}\left(\{ t_{k} \}\right)}$ in eq (\ref{defF}) into
the contribution from each genus;
\begin{equation}
  {{\cal F}\left(\{ t_{k} \}\right)}
  = \sum_{g=0}^{\infty} {\cal F}_{g}(\{ t_k \}),
 \label{genusexp}
\end{equation}
where ${\cal F}_g$ is the free energy of the $A_{N-1}$ topological string on
Riemann surface of genus $g$.  The Virasoro constraints
(\ref{Virasoro}) on ${\cal F}$ give us the nonlinear coupled conditions on
each ${\cal F}_g$;
\begin{equation}
 0=c_{g}^{(n)} ( {\cal F}_g, {\cal F}_{g-1}, \ldots, {\cal F}_0 ; \{t_k\} ),
\end{equation}
where $n\geq -1$ and $g\geq 0$.

For example, the constraints $c_{g=0,1}^{(n)} $ read as follows;
\begin{eqnarray}
0 &=& c_{g=0}^{(n)} \nonumber \\
  &=&
 \sum_{k\geq 1} k t_{k}
 \frac{\partial {\cal F}_0}{\partial t_{k+Nn}}
 + \frac{1}{2} \sum_{k+l=nN}
 \frac{\partial {\cal F}_0}{\partial t_{k}}
 \frac{\partial {\cal F}_0}{\partial t_{l}}
 +
 \frac{1}{2} \sum_{k+l=N} k l
  t_{k} t_{l}
 \delta_{n+1,0}
 \label{Virasorog=0}
 \\
0 &=& c_{g=1}^{(n)} \nonumber \\
  &=&
 \sum_{k\geq 1} k t_{k}
 \frac{\partial {\cal F}_0}{\partial t_{k+Nn}}
 + \sum_{k+l=nN}
 \frac{\partial {\cal F}_0}{\partial t_{k}}
 \frac{\partial {\cal F}_1}{\partial t_{l}}
 +
 \frac{1}{2}
 \sum_{k+l=nN}
 \frac{\partial^2 {\cal F}_0}{\partial t_{k} \partial t_{l}}
 +
 \frac{N^2-1}{24}
 \delta_{n,0}
 \label{Virasorog=1}
\end{eqnarray}

\vskip 3mm

Let us study the constraints (\ref{Virasorog=0}) on ${\cal F}_{g=0}$. It is
important to note that these constraints characterize a
$\tau$-function of $N$-reduced dispersionless KP system \cite{TT}
\cite{K}.

Dispersionless KP hierarchy is a ``quasi-classical'' ( $\hbar
\rightarrow 0$ ) limit of KP hierarchy in the following substitution;
\begin{equation}
  [ \hbar \partial_x, x ] = \hbar
  \longrightarrow
  \{ p, x\} = 1,
  \label{planck}
\end{equation}
where the Poisson bracket is defined by
\begin{equation}
  \{ f,g \} =
  \frac{\partial f}{\partial p}
  \frac{\partial g}{\partial x}
  -
  \frac{\partial f}{\partial x}
  \frac{\partial g}{\partial p}.
\end{equation}
In the paper \cite{TT} dispersionless KP hierarchy is formulated as an
analogue of Orlov's improved Lax formalism of KP hierarchy
\cite{Or}.The former is given by the following system of
equations;
\begin{eqnarray}
  && \frac{\partial L}{\partial t_{n}}
  = \{ B_{n}, L \}, \qquad (n\geq 1) \nonumber \\
  && \frac{\partial M}{\partial t_{n}}
  = \{ B_{n}, M \}, \qquad (n\geq 1) \nonumber \\
  && \{L,M\} = 1,
  \label{dlKP}
\end{eqnarray}
where $L$, $M$ are defined by
\begin{eqnarray}
  L &:=& p + \sum_{i=1}^{\infty} u_{i+1} (t) p^{-i}, \nonumber \\
  M &:=& L +
  \sum_{i=1}^{\infty} i t_{i} L^{i-1}
  + \sum_{i=1}^{\infty} v_{i+1}(t) L^{-i-1}
  \label{dlKPL}
\end{eqnarray}
and $B_n$ are given as truncation to positive powers of $p$;
\begin{equation}
  B_{n} := ( L^{n} )_{+}.
\end{equation}
The solution of the equations (\ref{dlKP}) is completely characterized
by a Riemann-Hilbert problem related to the area preserving
diffeomorphism on two plane \cite{TT}.  Especially the following
constraints specify an unique solution $(L,M)$ of the equations
(\ref{dlKP}) ;
\begin{equation}
   ( P^{n+1} Q^{m} )_{-} = 0 \qquad ( n \geq -1, m \geq 0 )
   \label{RH}
\end{equation}
where
\begin{equation}
  P = \frac{1}{N} L^{N}, \qquad Q=L^{1-N} M.
\end{equation}

In the case of $m=0$ the condition (\ref{RH}) are equivalent to
$(L^{N})_{-} = 0 $, that is, $N$-reduction of the system.  What about
the case of $m\geq 1$ ? Let us introduce ${\tau^{\rm d.l.KP}}(\{t_k\})$,
a $\tau$-function
of dispersionless KP hierarchy\cite{TT}, through the relation
\begin{equation}
  v_{n+1} = \frac{ \partial \log {\tau^{\rm d.l.KP}} }{\partial t_{n}}
  \qquad (n \geq 1).
  \label{dlTAU}
\end{equation}
Then we can show that in the case of $m=1$ the constraints (\ref{RH})
on $\log {\tau^{\rm d.l.KP}}$ precisely coincide with
$c_{g=0}^{(n)}=0 \quad (n\geq-1)$ constraints on ${\cal F}_{g=0}$
given in eq (\ref{Virasorog=0}). Hence the following relation holds;
\begin{equation}
  {\cal F}_{g=0} = \log {\tau^{\rm d.l.KP}},
  \label{Fg=0}
\end{equation}
where the $\tau$-function ${\tau^{\rm d.l.KP}}$ is the solution of (\ref{RH}).

At this stage let us give some comments.

1. The condition (\ref{RH}) in the case of $m\geq2$ are equal to
those on ${\cal F}_{g=0}$ obtained by the genus expansion
(\ref{genusexp}) from the $W$-constraints which ${\cal F}$ defined
in eq (\ref{GKM}) satisfies \cite{Morozov}.

2. By comparing two equations (\ref{genus}) and (\ref{planck}),
the genus expansion is nothing but the $\hbar$ expansion;
\begin{equation}
  {\cal F} = \sum_{g=0}^{\infty} \hbar^{\chi_{g}} {\cal F}_{g}
\end{equation}
where $\chi_{g} = 2 (g-1)$ is the Euler characteristic
of the genus $g$ Riemann surface.

3. By using the technique of dispersionless KP hierarchy,
we can show that
$ c_{g=0}^{(n=-1)} $ implies the  recursion relation
on punctured sphere \cite{DW}\footnote{
The derivation of the above recursion relation in the framework
of dispersionless KP hierarchy was first given by Prof. K. Takasaki.};
\begin{equation}
  \frac{\partial^{3} {\cal F}_{0}}{
  \partial t_m
  \partial t_i
  \partial t_j
  }
 =
  m \sum_{l=1}^{N-1}
  \frac{1}{l(N-l)}
  \frac{\partial^{2} {\cal F}_{0}}{
  \partial t_{m-N}
  \partial t_l
  }
  \frac{\partial^{3} {\cal F}_{0}}{
  \partial t_{N-l}
  \partial t_i
  \partial t_j
  },
  \label{recg=0}
\end{equation}
where $m \geq N+1$.

4. By the identification  (\ref{Fg=0}), we can check that the
following form of ${\cal F}_{g=1}$ is the solution of
$ c_{g=1}^{(n)} = 0 $ (see eq (\ref{Virasorog=1}));
\begin{equation}
  {\cal F}_{g=1} = \frac{1}{24} \log \left[
  {\rm det~} \left(
  \frac{\partial^{3} {\cal F}_{0}}{
  \partial t_{1}
  \partial t_i
  \partial t_j
  }
  \right)_{1 \leq i,j \leq N-1}
  \right],
  \label{Fg=1}
\end{equation}
which is exactly the conjectured formula (up to an additive constant)
given in \cite{DW}.

\vskip 3mm
In order to proceed further, let us change our perspective.  We shall
study the perturbed $A_{N-1}$ topological minimal matter on the sphere
in the presence of constant background of gravity descendants.  Namely
we treat the gravitational descendants as external fields with some
fixed strength.  Let $\alpha =(\alpha_{N+1},\alpha_{N+2},\cdots)$ be a
sequence of constants where only a finite number of $\alpha_i$ are non
zero.  We introduce the following $(N-1)$ dimensional subspace
$V_{\alpha }$ of the infinite dimensional parameter space (full phase
space);
\begin{eqnarray}
V_{\alpha} =
\{ (t_1,t_2, \cdots,t_{N-1},
 t_{N+1}=\alpha_{N+1},
 t_{N+2}=\alpha_{N+2}, \cdots ) \},
\label{small}
\end{eqnarray}
where $t_{k}(k \geq N+1)$ are constrained to the value $\alpha_{k}$.
The conventional small phase space of the perturbed $A_{N-1}$
topological minimal matter on the sphere \cite{DVV2} is given by
$V_{\alpha_{0}}$ with $\alpha_0:=(\frac{1}{N+1},0,0,\cdots)$ \cite{TT,K}.

We shall take the following Legendre transformed variables;
\begin{eqnarray}
q^{(\alpha)}_{i+1}:=
\left. - \frac{1}{i}
\frac{\partial^{2} {\cal F}_{g=0}}{\partial t_1 \partial t_i}
\right|_{V_{\alpha}}
{}~~~~(1 \leq i \leq N-1).
\label{qalpha}
\end{eqnarray}
Notice that, via $0=C_{g=0}^{(n=-1)}$ (\ref{Virasorog=0}), we can
write down $q_{i+1}^{(\alpha)}$ perturbatively in terms of
$t_{1},\cdots ,t_{N-1}$ .  Especially, on $V_{\alpha_0}$, $q_{i+1}:=
q_{i+1}^{(\alpha_0)}$ get the following forms;
\begin{eqnarray}
q_{i+1}=(N-i)t_{N-i}~~~ (1\leq i \leq N-1),
\label{q}
\end{eqnarray}
that is, $q^{(\alpha)}_{i+1}$ reduce to the time variables .  So we
may take $q^{(\alpha)}$ as the coordinates of the subspace $\alpha
\sim \alpha_0$ .

What is the relation between the physics on $V_{\alpha_0}$ and on
$V_{\alpha \neq \alpha_{0}}$ ?  Because of (\ref{qalpha}) and
(\ref{q}) it may be measured by the Jacobian associated with the
change of coordinates $q^{(\alpha_0)} \rightarrow q^{(\alpha)}$,
$\mbox{det}\left( \frac{\partial q_{i+1}^{(\alpha)}}{\partial t_{j}}
\right)_{1 \leq i,j \leq N-1}$,
that is,
\begin{eqnarray}
\left.
\mbox{det}
\left(
\frac{\partial^3 {\cal F}_{g=0}}
     {\partial t_1 \partial t_i \partial t_j}
\right)_{1 \leq i,j \leq N-1}\right|_{V_{\alpha}}
=
\left.
e^{24 {\cal F}_{g=1}}
\right|_{V_{\alpha}},
\label{Jacobian}
\end{eqnarray}
where we use the relation (\ref{Fg=1}).  The metric $\eta$ of the
perturbed $A_{N-1}$ topological minimal matter is given by
$\eta_{ij}:=\left.\frac{\partial^3 {\cal F}_{g=0}}
{\partial t_1 \partial t_i \partial t_j}\right|_{V_{\alpha_0}}$ and it
is deformed as $\alpha$ changes from $\alpha_0$ into $\alpha_1$.  Let
us assume that the R.H.S of (\ref{Jacobian}) vanishes at
$V_{\alpha_1}$.  Then, finally at $\alpha=\alpha_1$, the deformed
metric $\eta_{ij}$ becomes degenerate.  This phenomena can be
described as ``gravitational phase transition'' of the perturbed
$A_{N-1}$ topological minimal matter, which may be interpreted as a
simple example of ``BRST symmetry breaking'' introduced in \cite{W2}.
We also note, because of the relation (\ref{Jacobian}), the
$\tau$-function (\ref{GKM}) becomes zero at $\alpha_{1}$;
\begin{eqnarray}
\left. e^{{\cal F}}\right|_{V_{\alpha_1}}=0.
\end{eqnarray}
This property reflects the analyticity of the wave function
\cite{SHI} associated with $e^{{\cal F}}$ (\ref{GKM}),
which predicts the above scenario.

It may be helpful to give the following remarks.

In topological field theories,the physical Hilbert spaces are finite
dimensional, and it appears strange to speak of their ``phase
transition'' which is, as is often explained, the cooperative
phenomena of infinitely many degrees of freedom.  However, we all know
that mean field theories, although correlation effect are totally
neglected, can explain the phase transition in a very simple manner.
Let us take Ising model, for example.  The free energy of the system
is a function of the external magnetic field and the temperature.  The
convexity of the free energy enables us to express the free energy in
terms of the magnetization.  The phase transition occurs at the point
where this Legendre transformation becomes singular.  This is exactly
what is happening in the topological field theory coupled to
topological gravity.  In the case of minimal topological matter theory
without gravity, free energy is a simple polynomial and no phase
transition occurs. In this sense, it may be called ``gravitational
phase transition''.

\vskip 3mm

The perturbed $A_{N-1}$ topological minimal matter on $V_{\alpha_0}$
( on the sphere )
is known to be deeply related with a versal deformation of a simple
singularity of type $A_{N-1}$ \cite{DVV2}.  Let us investigate this
relation in our framework.  Up to now, the perturbed $A_{N-1}$
topological minimal matter on $V_{\alpha}$ is described using the
terminology of $N$-reduced dispersionless KP system.  In order to step
into deformations of $A_{N-1}$ simple singularity, we should make
clear the correspondence between these two objects.  For this purpose
we introduce a concept of hierarchy into simple singularities
of type $A$ \cite{IN}, which will build a bridge between them.

Consider a formal Laurent series;
\begin{eqnarray}
L(p):=
p+ \sum_{i \geq 1}u_{i+1}p^{-i}~~,
\label{fracL}
\end{eqnarray}
where $u=(u_2.u_3, \cdots )$ is a sequence of countably many
variables.  Then one can find a unique Laurent series ;
\begin{eqnarray}
p(L):=L +\sum_{i \geq 1}q_{i+1}L^{-i}~~,
\label{fracP}
\end{eqnarray}
such that $p \circ L(y)=y$ and $L \circ p(y)=y$.  Hence we obtain a
new sequence $q=(q_2,q_3, \cdots)$ of countably many variables.  It
should be noted that any time evolution such as (\ref{dlKPL}) is not
yet assumed in the Laurent series $L$ in eq (\ref{fracL}).  Define the
following set of polynomials ;
\begin{eqnarray}
\phi_{i}(p)
:=\frac{1}{i+1}\partial_{p}(L^{N+1})_{+}~~~~~(i \geq 0),
\label{phi}
\end{eqnarray}
which satisfies ;
\begin{eqnarray}
\mbox{res}\left\{
\frac{\phi_i \phi_j}{\phi_k}
dp \right\}=
\delta_{i+j,k-1}~~~(0 \leq i,j \leq k-1).
\end{eqnarray}
Introduce their generating function by
\begin{eqnarray}
\phi(p;\lambda):=
\sum_{i \geq 0} \phi_i(p)\lambda^{i},
\end{eqnarray}
and then we can see that it satisfies the fundamental relation
\cite{IN};
\begin{eqnarray}
\lefteqn{ \lambda \mu \mbox{res}
\left\{
\frac{\phi(\lambda)\phi(\mu)L}{\partial_{p}L}
dp \right\}} \nonumber \\
&=&
(\lambda \partial_{\lambda}+\mu \partial_{\mu})
\log \left( 1-\lambda \mu
        \sum_{n \geq 2}q_n
        \frac{\lambda^{n-1}-\mu^{n-1}}{\lambda-\mu} \right).
\label{fracF}
\end{eqnarray}

On the other hand, in dispersionless KP hierarchy,
any $\tau$-function (\ref{dlTAU}) enjoys the same property;
\begin{eqnarray}
&&
 \sum_{n,m \geq 1}\lambda^n \mu^m
\frac{\partial_{t_n}}{n}
\frac{\partial_{t_m}}{m}
\log {\tau^{\rm d.l.KP}} \nonumber \\
&&~~~~~~
= \log \left( 1+\lambda \mu
        \sum_{n \geq 1}
        \frac{\lambda^{n}-\mu^{n}}{\lambda-\mu}
          \frac{\partial_{t_n}}{n}
             \partial_{t_1}
                 \log {\tau^{\rm d.l.KP}}\right),
\label{SDR}
\end{eqnarray}
which is the ``quasi-classical'' limit of the corresponding relation
in KP hierarchy.  By comparing (\ref{fracF}) with (\ref{SDR}) we are
naturally lead to establish the universal correspondences ;
\begin{eqnarray}
&&q_{n+1}=-\frac{1}{n}
         \partial_{t_n}\partial_{t_1}
             \log {\tau^{\rm d.l.KP}}
\label{flat}
\\
&&\mbox{res}\left\{
\frac{\phi_{m-1}\phi_{n-1}L}{\partial_{p}L}
dp \right\}
=
\frac{mn}{m+n}
\partial_{t_n}\partial_{t_m}
   \log {\tau^{\rm d.l.KP}}
\label{3pt}
\end{eqnarray}
Notice that these correspondences hold for arbitrary $\tau$ function
of dispersionless KP hierarchy (independent of $N$).  Any
$u=(u_1,u_2,\cdots)$ in (\ref{fracL}) can be parameterized by
$t=(t_1,t_2,\cdots)$ (KP-times) with choosing a $\tau$ function which
satisfies (\ref{flat}),(\ref{3pt}).  Of course we can say it in the
reverse order. That is, we can get a $\tau$ function of dispersionless
KP hierarchy by imposing the conditions (\ref{flat}),(\ref{3pt}) on
$u$ and $t$.

With the above correspondences (\ref{flat}),(\ref{3pt}) we can easily
clarify the relation between the perturbed $A_{N-1}$ topological
minimal matter on $V_{\alpha_0}$ and a versal deformation of a simple
singularity of type $A_{N-1}$.  Impose the following condition on
$L(p)$ (\ref{fracL});
\begin{eqnarray}
L^N &=& \left( L^N \right)_+   \nonumber \\
    &=&
     p^N +a_2p^{N-2}+ \cdots +a_{N-1}p+ a_N,
\label{singular}
\end{eqnarray}
which can be regarded as a versal deformation of the singularity
$p^N=0$.  With this constraint $q=(q_2,q_3,\cdots)$ in (\ref{fracP})
are weighted homogeneous polynomials with respect to $a_2, \cdots,
a_N$.  Especially $y=(y_2, \cdots, y_{N})$, the flat coordinates
\cite{SSY} of this versal deformation, can be taken as \cite{IN};
\begin{eqnarray}
 y_i= -N q_i~~~~( 2 \leq i \leq N).
\label{flatAN-1}
\end{eqnarray}
On $V_{\alpha_0}$, applying the correspondence (\ref{flat}) to
(\ref{Fg=0}), this flat coordinate system can be rephrased as;
\begin{eqnarray}
  y_i=-N(N+1-i)t_{N+1-i}.
\label{flatAN-12}
\end{eqnarray}
Notice that, by combining the relations (\ref{flat}),(\ref{flatAN-1}),
we can describe the perturbed $A_{N-1}$ minimal matter on $V_{\alpha
\neq \alpha_{0}}$ in terms of singularity theory \cite{noumiseminar}.

In general, a flat coordinate system was introduced as a special
coordinate system in order to study the Gauss-Manin system associated
with a versal deformation of an isolated singularity \cite{Saito}.
A detailed study of these Gauss-Manin systems is given in \cite{N}. What is
the relation between the $A_{N-1}$ type Gauss-Manin system and the
$A_{N-1}$ topological string ? It may shed some insights on
topological string theory.  Via the correspondences (\ref{flat}),
(\ref{3pt}) we may extend the $A_{N-1}$ Gauss-Manin system in terms of
topological string.  The discriminant $\Delta$ of (\ref{singular});
\begin{equation}
  \Delta := {\rm det~}
  \left(
  {\rm res~}
  \left\{
   \frac{ \phi_{i} \phi_{N-2-j} L}{ \partial_{p} L } dp
  \right\}
  \right)_{0\leq i,j\leq N-2}
\end{equation}
satisfies the equations \cite{IN};
\begin{equation}
  \theta_{k} ( \Delta )  =
  \frac{\partial \hat{\tau}}{\partial y_{k}}  \Delta
  \qquad ( 2 \leq k \leq N),
\end{equation}
where $\theta_{k} ~ (2 \leq k \leq N)$ are the logarithmic vector fields
\cite{Saito} given by ;
\begin{equation}
  \theta_{k} := \sum_{i=2}^{N}
  {\rm res~}
  \left\{
   \frac{ \phi_{N-i} \phi_{i-2} L}{ \partial_{p} L } dp
  \right\}
  \frac{\partial}{\partial y_{k}},
\end{equation}
and $ \hat{\tau} $ is
\begin{equation}
  \hat{\tau} := \sum_{i=0}^{N-2}
  {\rm res~}
  \left\{
   \frac{ \phi_{i} \phi_{N-2-i} L}{ \partial_{p} L } dp
  \right\}.
\end{equation}
We expect that these quantities have meaning in the
$A_{N-1}$ topological string theory through the
correspondences
(\ref{flat}), (\ref{3pt}) and (\ref{flatAN-1}).

\vskip 3mm

We would like to thank Prof. K. Takasaki,
Prof. T. Shiota, and Dr. A. Nagai for useful discussions.


\begin{thebibliography}{99}

\bibitem{Lossev}
 A. Lossev,
``Descendants constructed from matter field in topological
Landau-Ginzburg theories coupled to topological gravity'',
IPI-MINN-92-40-T (1992).
\bibitem{EKYY}
 T. Eguchi, H. Kanno, Y. Yamada and S. -K. Yang, Phys. Lett. B305 (1993) 235.
\bibitem{Kon}
 M. L. Kontsevich, Comm. Math. Phys. 147 (1992) 1.
\bibitem{Morozov}
 S. Kharchev, A. Marshakov, A. Mironov, A. Morozov
 and A. Zabrodin, Nucl. Phys. B380 (1991) 181;
 C. Itzkson and J. B. Zuber, Int. Journ. Mod. Phys. A7 (1992) 5661;
 S. Kharchev, A. Marshakov, A. Mironov and A. Morozov,
 Mod. Phys. Lett. A8 (1993) 1047.
\bibitem{noumiseminar}
 A. Kato, T. Nakatsu, M. Noumi and T. Takabe, in preparation.
\bibitem{witten1}
 E. Witten, Nucl. Phys. B371 (1992) 191.
\bibitem{DVV}
 R. Dijkgraaf, E. Verlinde and H. Verlinde, Nucl. Phys. B348 (1991) 435;
 M. Fukuma, H. Kawai and R. Nakayama, Int. Journ. Mod. Phys. A6 (1992) 1385.
\bibitem{TT}
 K. Takasaki and T. Takebe, Int. Journ. Mod. Phys. A7. suppl. 1B (1992) 889;
``Quasiclassical Limit of KP Hierarchy, $W$-Symmetries and Free
Fermions'', preprint KUCP-0050/92, hep-th/9207081.
\bibitem{K}
 I. M. Krichever, Comm. Math. Phys. 143 (1992) 415.
\bibitem{Or}
 P. G. Grinevich and A. Yu. Orlov, in ``Problems of Modern
 Quantum Field Theory'', Springer-Verlag (1989).
\bibitem{DW}
 R. Dijkgraaf and E. Witten, Nucl. Phys. B342 (1992) 486.
\bibitem{DVV2}
 R. Dijkgraaf, E. Verlinde and H. Verlinde, Nucl. Phys. B352 (1991) 59.
\bibitem{W2}
 E. Witten, Comm. Math. Phys. 118 (1988) 411.
\bibitem{SHI}
 T. Shiota,  private communication.
\bibitem{IN}
 S. Ishiura and M. Noumi, Proc. Japan. Acad. 58 (1982) 13, 62;
 ``Gauss-Manin systems of type $A$'' RIMS koukyuroku 459 (1982) 29
 (in Japanese).
\bibitem{SSY}
 K. Saito, T. Yano and J. Sekiguchi, Comm. Algebra. 8 (1980) 373.
\bibitem{Saito}
 K. Saito, J. Fac. Sci. Uni. Tokyo. Sec. IA. 28 (1982) 775.
\bibitem{N}
 M. Noumi, Tokyo. J. Math. 7 (1984) 1.
\end{thebibliography}
\end{document}